\documentclass[aps, preprint, eqsecnum, showpacs, epsfig]{revtex4}
\usepackage{graphics}
\usepackage{graphicx}
\usepackage{epsfig}

\def \be {\begin{equation}}
\def \ee {\end{equation}}
\def \ba {\begin{eqnarray}}
\def \ea {\end{eqnarray}}
\def \bm {\begin{displaymath}}
\def \em {\end{displaymath}}
\def \br {{\bf r}}
\def \bom {{\bf \Omega}}

\begin{document}
\title{Pair Correlation Functions and a Free-Energy Functional for the Nematic Phase}
\author{Pankaj Mishra, Swarn Lata Singh, Jokhan Ram and Yashwant Singh}
\affiliation{Department of Physics, Banaras Hindu University, 
Varanasi-221 005,
India}
\date{\today}
\begin{abstract}
In this paper we have presented the calculation of pair correlation functions in a nematic phase for a model 
of spherical particles with the long-range anisotropic interaction from the mean spherical approximation(MSA)
and the Percus-Yevick (PY) integral equation theories. The results found from the 
MSA theory have been compared with those found analytically by Holovko and Sokolovska (J. Mol. Liq. 
$\bf 82$, 161(1999)). A free energy functional which involves both the symmetry conserving and symmetry broken 
parts of the direct pair correlation function has been used to study the properties
of the nematic phase. We have also examined the possibility of constructing a free 
energy functional with the direct pair correlation function which includes only the principal order parameter 
of the ordered phase
and found that the resulting functional gives results 
that are in good agreement with the original functional. The isotropic-nematic transition has been located using
the grand thermodynamic potential. The PY theory has been found to give nematic phase with pair 
correlation function harmonic coefficients having all the desired features. In a nematic phase 
the harmonic coefficient of the total pair correlation function
$h({\bf x_1},{\bf x_2})$ connected with the correlations of the 
director transverse fluctuations should develop a long-range tail. This
feature has been found in both the MSA and PY theories. 

\end{abstract}
\pacs{71.15.Mb, 64.70.Md, 61.30.Cz}
\maketitle
\section{Introduction}
The distribution of molecules in a classical system can adequately be described by one and two-particle 
density distributions. The one particle density distribution, $\rho({\bf x})$ defined as
\be
\rho({\bf x})=\rho({\br}, \bom)=<\sum_{i=1}\delta(\br - \br_{i})\delta(\bom-\bom_{i})>
\ee
where ${\bf x}_i$ indicates both position $\br_{i}$ and orientation $\bom_i$ of $i^{th}$ molecule,  the angular 
bracket represents the ensemble average and $\delta$ the Dirac function, is constant 
independent of position and orientation for an isotropic fluid but contains most of the structural 
informations of ordered phases like crystalline solids and liquid crystals. 
The two-particle density distribution $\rho({\bf x}_{1}, {\bf x}_{2})$ which gives probability 
of finding simultaneously a molecule in volume element $d\br_{1}$$d\bom_{1}$ 
centered at $(\br_{1},\bom_{1})$ and a second molecule in volume element 
$d\br_{2}$$d\bom_{2}$ centered at  $(\br_{2},\bom_{2})$ is defined as
\be
\rho({\bf x}_1, {\bf x}_2)=\rho(\br_{1}, \bom_{1}, \br_{2}, \bom_{2})=
<\sum_{i\neq j}\delta(\br_{1} - \br_{i})\delta(\bom_{1} -\bom_{i})\delta(\br_{2} - \br_{j})\delta(\bom_{2} -\bom_{j})>
\ee
The pair correlation function $g({\bf x}_1, {\bf x}_2)$ is related to $\rho({\bf x}_1, {\bf x}_2)$ by the relation 
\be
g({\bf x}_1, {\bf x}_2)=\frac{\rho({\bf x}_1, {\bf x}_2)}{\rho({\bf x}_1) \rho({\bf x}_2)}
\ee
Since in an isotropic fluid $\rho({\bf x}_1)= \rho({\bf x}_2)= \rho_{l}= \frac{<N>}{V}$ where $<N>$ is the 
average number of molecules in volume V,
\be
\rho^{2}_{l} g(\br, \bom_{1}, \bom_{2})=\rho(\br, \bom_{1}, \bom_{2})
\ee
where $\br=(\br_2-\br_1)$. In the isotropic fluid $g({\bf x}_1, {\bf x}_ 2)$ depends only on inter particle 
distance $|\br_2-\br_1|=r$, orientation of molecules with respect to each other and on the direction 
of vector $\br $ ($\hat \br=\frac{\br }{r}$ is a unit vector along r). These simplifications are due to 
homogeneity which implies continuous translational symmetry and isotropy which implies continuous rotational symmetry. Such simplifications do not generally occur in ordered phases.

The pair correlation functions as a function of intermolecular separations and orientations at a 
given temperature and pressure can be found either by computer simulation[1-5] or by simultaneous 
solution of an integral equation, the Ornstein-Zernike (OZ) equation,

\be
h({\bf x}_1,{\bf x}_2)=c({\bf x}_1, {\bf x}_2)+\int c({\bf x}_1, {\bf x}_3)\rho({\bf x}_3)h({\bf x}_3, {\bf x}_2) d{\bf x}_3
\ee
where $d{\bf x}_3=d\br_{3}d\bom_{3}$ and $h({\bf x}_1, {\bf x}_2)(=g({\bf x}_1, {\bf x}_2)-1$) 
and $c({\bf x}_1, {\bf x}_2)$ are respectively, the total and direct pair correlation functions, and an 
algebraic closure relation which relates the correlation functions to the pair potential. 
Well known approximations to the closure relation are the hypernetted-chain relation, 
the Percus-Yevick (PY) relation and the mean spherical approximation (MSA) [6]. These integral 
equation theories have been quite successful in describing the structure and thermodynamic properties
 of isotropic fluids [7-11]. However, their application to ordered phases which can be regarded 
as inhomogeneous, have so far been very limited [12-15], though no feature of the theory inherently 
prevents them from being used to describe the structure of ordered phases. One of the problems 
that arises in the case of ordered phases is the appearance of $\rho(\bf r, \bf\Omega)$ in the 
OZ equation (see Eq.(1.5)). This implies that in contrast to the isotropic case where we needed 
only two relations, namely the OZ equation and a closure relation, an additional relation corresponding 
to single particle distribution connecting to pair correlation function is needed to solve 
the ensuing equations self consistently. 

In this paper we take nematic in which molecules are aligned on the average along a particular but 
arbitrary direction while the translational degrees of freedom remain disordered as in an 
isotropic phase, as an example of an ordered phase. At the isotropic-nematic transition 
the isotropy of the space is spontaneously broken and as a consequence, the correlations 
in the distribution of molecules lose their rotational invariance. The change from 
isotropic fluid to nematic state in the absence of external field involves collective 
fluctuations, which develops orientational wave excitations known as Goldstone modes[16]. 
This leads to the divergence of the corresponding harmonics of the total pair correlation 
function $h({\bf x}_1,{\bf x}_2)$ in the limit of  zero wave vector. By computer simulation of a system of 
ellipsoids Phoung and Schmid [13] have evaluated the effect of breaking of rotational symmetry 
on pair correlation functions and showed that in a nematic phase there are two qualitatively 
different contributions; one that preserves rotational invariance and the other that breaks 
it and vanishes in the isotropic phase.

Holovko and Sokolovska [14] have used the MSA closure 
relation and the Lovett equation [17] (see Eq.(3.17))which relates one particle density to pair correlation 
function to solve analytically the OZ equation for a model of spherical particles with the 
long range anisotropic interaction (see Eq.(2.1)) in a nematic phase. However, when  Phoung 
and Schmid [13] used the PY closure and the Lovett equation and solved the OZ equation numerically 
for a system of soft ellipsoids, nematic phase was not found and for this the PY closure was blamed.
Zhong and Petschek [18] have analyzed the diagrammatic expansion of the direct correlation function and
concluded that the PY closure can not reproduce the Goldstone modes in the general case of spontaneous partial
ordering.

Recently we [19] used the PY closure and solved numerically the OZ equation for a system of 
elongated rigid molecules interacting via the Gay-Berne potential [20] and showed that the PY 
closure gives  nematic phase with the pair correlation function harmonic coefficients 
having features similar to those found by computer simulation [13] and by analytical solution [14].
Instead of using a closure relation for $\rho(\br, \bom)$ we expressed it in terms of order 
parameters and solved the resulting equation for values of order parameters ranging from zero to 
some maximum value. Non-zero values of order parameters break the symmetry of isotropic phase and 
the degree of symmetry breaking is given by the values of the order parameters. 
Using these correlation functions we constructed a free energy functional and used it 
to determine the value of order parameters in the nematic phase by minimizing it. 
Once the values of order parameters are known the pair correlation functions in the nematic phase
are obtained from the known results.

In this paper we extend our method to calculate the pair correlation functions in nematic phase 
using the MSA and PY closure relations for a system the molecules of which interact via a pair 
potential considered in ref. [14]. This allows us to compare our results for the MSA with those 
found analytically and therefore to test the accuracy of our method. The PY relation is shown 
to give nematic phase with all the expected features. The paper is organized as follows: In Sec.II
we describe the MSA and PY integral equation theories and give a brief account of computational procedure.
 In Sec. III we construct a free energy functional of an inhomogeneous system that contains both symmetry
conserved and symmetry broken parts of the direct pair correlation function. The isotropic-nematic transition
point and freezing parameters are calculated in Sec IV. The paper ends with discussions given in Sec. V.
 
\section{Correlation Functions}

The pair potential used by Holovko and Sokolovska [14] in their analytical solution of the OZ and 
Lovett equation with MSA closure has the form 
\be
v({\bf x}_1,{\bf x}_2)= v_{hs}(r)+v_0(r)+v_{2}(r, \Omega_1, \Omega_2)
\ee
where $v_{hs}$ is the hard sphere potential. 
\ba
v_{hs}(r)& = &\infty {\hspace{1cm}} r<{\sigma} \\
& = & 0  {\hspace{1.2cm}} r>{\sigma}
\ea
The long-range attraction has isotropic part
\be
v_0(r)=-a_0(z_0\sigma)^{2}\frac{\exp(-z_0 r)}{r/\sigma}
\ee
and the anisotropic part
\be
v_{2}(r, \Omega_1, \Omega_2)=-a_2(z_2\sigma)^{2}\frac{\exp(-z_2 r)}{r/\sigma}P_{2}(\cos\Omega_{12})
\ee
where $P_{2}(\cos\Omega_{12})$ is the second order Legendre polynomial of relative molecular orientations.
This model potential is independent of orientation of the intermolecular separation vector {\br}. This fact
limits the number of harmonic coefficients that appear in the spherical harmonic expansion of pair
 correlation functions.

We choose a coordinate frame with it's z-axis in the direction of the director $\hat n$(director frame).
The director $\hat n$ is a unit vector along the direction of alignment of molecules.
All orientation dependent functions are expanded in spherical harmonics $Y_{lm}(\Omega)$[6]. This yields 
(for uniaxial nematic phase of axially symmetric molecules) [21]
\ba
\rho(\br, \bom)&=& \rho f(\Omega) \nonumber \\
&=& \frac{\rho}{\sqrt{4\pi}}\sum_{l({\rm even})}f_{l}Y_{l0}(\Omega)
\ea
 
where $f_l=\sqrt{(2 l + 1)}P_l$ and $f_0=1$. $P_l$ for $l\neq 0$ are order parameters; their values are 
zero in the isotropic phase and nonzero in the nematic phase. For two particle functions one has [22]  
\ba
\psi(r,{\Omega_1}, {\Omega_2}) = 
\sum_{l_1 l_2 l m_1 m_2 m}& &\psi_{l_1 l_2 l m_1 m_2 m}(r)
 Y_{l_1m_1}(\Omega_1) Y_{l_2m_2}(\Omega_2) Y_{lm}^{*}({\hat{\bf r}})
\ea

where $\psi$ stands for $h$ or $c$ or $v$. In uniaxial nematic phases, only real coefficients with $m_1+m_2-m=0$
and even $l_1+l_2+l$ enter in the expansion. Since the molecules in the model
system under consideration have axial symmetry, every single $l$ is even as well. Because, in isotropic 
phase $h$ and $c$ preserve the rotational symmetry, for them
\be
\psi_{l_1 l_2 l m_1 m_2 m}(r)=\psi_{l_1l_2l}(r) C_g(l_1 l_2 l m_1 m_2 m)
\ee

where $C_g$ is the Clebsch-Gordan(CG) coefficient. The absence of the CG coefficients in Eq.(2.7) when $\psi$
represents  pair correlation functions of nematic phase, removes the restriction $|l_1-l_2|\leq l \leq l_1+l_2$
on the values of the index $l$. As a consequence, coefficients such as $\psi_{200000}(r)$ and $\psi_{020000}(r)$
are nonzero in nematic whereas they do not survive in the isotropic case. The emergence of these harmonic 
coefficients are due to symmetry breaking.

To solve the OZ equation it is advisable to use the Fourier representation. The expansion coefficients 
$\psi_{l_1 l_2 l m_1 m_2 m}(r)$ are related to their counterparts in Fourier space by the Hankel transform
\ba
{\hat\psi}_{l_1 l_2 l m_1 m_2 m}(k)&=&4\pi i^{l}\int_{0}^{\infty}dr r^{2} j_{l}(kr)\psi_{l_1 l_2 l m_1 m_2 m}(r),\\
{\hat\psi}_{l_1 l_2 l m_1 m_2 m}(r)&=&\frac{4\pi(-i)^{l}}{{(2\pi)}^{3}}
\int_{0}^{\infty}dk k^{2} j_{l}(kr)\psi_{l_1 l_2 l m_1 m_2 m}(k)
\ea
where $j_l(kr)$ is the spherical Bessel function.

Using Eqs(2.9)-(2.10) and the spherical harmonic expansion for the correlation functions (Eqs.(2.6) and (2.7)), OZ
equation reduces in the k-space to the form
\ba
h_{l_1l_2lm_1m_2m}(k)-c_{l_1l_2lm_1m_2m}(k)&=& \gamma_{l_1l_2lm_1m_2m}(k)\nonumber \\
&=&\frac{\rho^*}{\sqrt{4\pi}}
\sum_{I} f_{l_{3}^{'''}}c_{l_1 l_{3}^{'}l^{'}m_1 m_{3}^{'}m^{'}}(k)h_{l_{3}^{''}l_2 l^{''}m_{3}^{''}m_{2}m^{''}}(k) \nonumber \\
&&\Gamma_{m_{3}^{'}m_{3}^{''}0}^{l_{3}^{'}l_{3}^{''}l_{3}^{'''}}
\hspace{0.3cm}\Gamma_{m^{'}m^{''}m}^{l^{'}l^{''}l}
\ea
where $\rho^*=\rho \sigma^3$ , the symbol $'I'$ indicates the collection of nine indices
$l_{3}^{'}, l_{3}^{''}, l_{3}^{'''}, l^{'}, l^{''}, m_{3}^{'}$, $m_{3}^{''}, m^{'}, m^{''}$ and notation
\ba
\Gamma_{m_1m_2m}^{l_1l_2l}&=& \int d\Omega Y_{lm}^{*}(\Omega)Y_{l_1m_1}(\Omega)Y_{l_2m_2}(\Omega)\nonumber \\
&=& \sqrt{\frac{(2l_1+1)(2l_2+1)}{4\pi(2l+1)}}C_{g}(l_1 l_2 l 0 0 0) C_{g}(l_1 l_2 l m_1 m_2 m)
\ea

Since the pair potential of Eq(2.1) is independent of orientation of the intermolecular separation vector 
$\bf r$, the harmonic coefficients that survive in the expansion of pair correlation functions have
$l=m=0$ and $m_1=-m_2$. This allows 
notational simplification from six indices to three. We therefore rewrite Eq. (2.11) as 
\ba
\gamma_{l_1l_20m-m0}(k)&\equiv&\gamma_{l_1l_2m}(k) \nonumber \\
&=& \frac{\rho^*}{4\pi}\sum_{l_{3}^{'}l_{3}^{''}}c_{l_1l_{3}^{'}m}(k)h_{l_{3}^{''}l_2m}(k)
\sum_{L}{P}_L\sqrt{2L+1}\Gamma_{{\underline m}m0}^{l_{3}^{'}l_{3}^{''}L}
\ea
where ${\underline m}=-m$.

Since there is no summation over index $m$ on the right hand side of Eq.(2.13), the OZ equation for 
harmonics with different values of $m$ decouple. The equation corresponding to the isotropic 
case is found by putting $P_0=1$ and $P_{L\neq 0}=0$ in Eq(2.13). 
\begin{figure}[ht]
\begin{center}
\vspace{1.2cm}
\includegraphics[height=3.0in,width=5.0in]{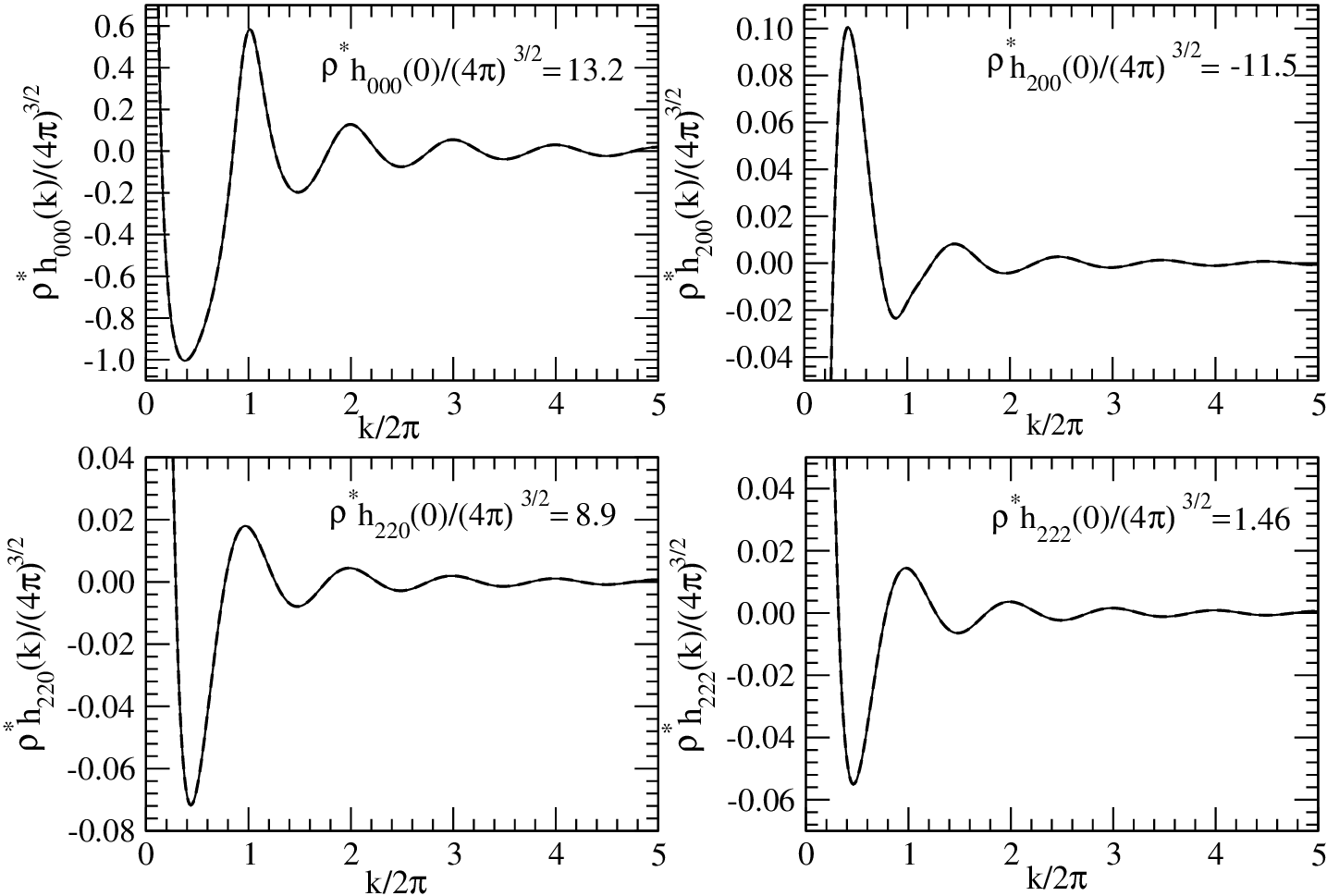}
\caption{Comparison of some of the harmonics of the total pair correlation function in the Fourier space
for the nematic phase ($\beta a_2=1, \beta a_0=0.1, z_0\sigma=z_2\sigma=1$, $\eta=0.315$, $P_2=0.63, P_4=0.27$)
 obtained by the analytical
solution[14](dashed line) with the results found using numerical method (full line) for the MSA. The two
curves are indistinguishable at the scale of the figure.}
\end{center}
\end{figure}
                                                                                                                             
\begin{figure}[h]
\begin{center}
\vspace{1.2cm}
\includegraphics[height=3.0in,width=5.0in]{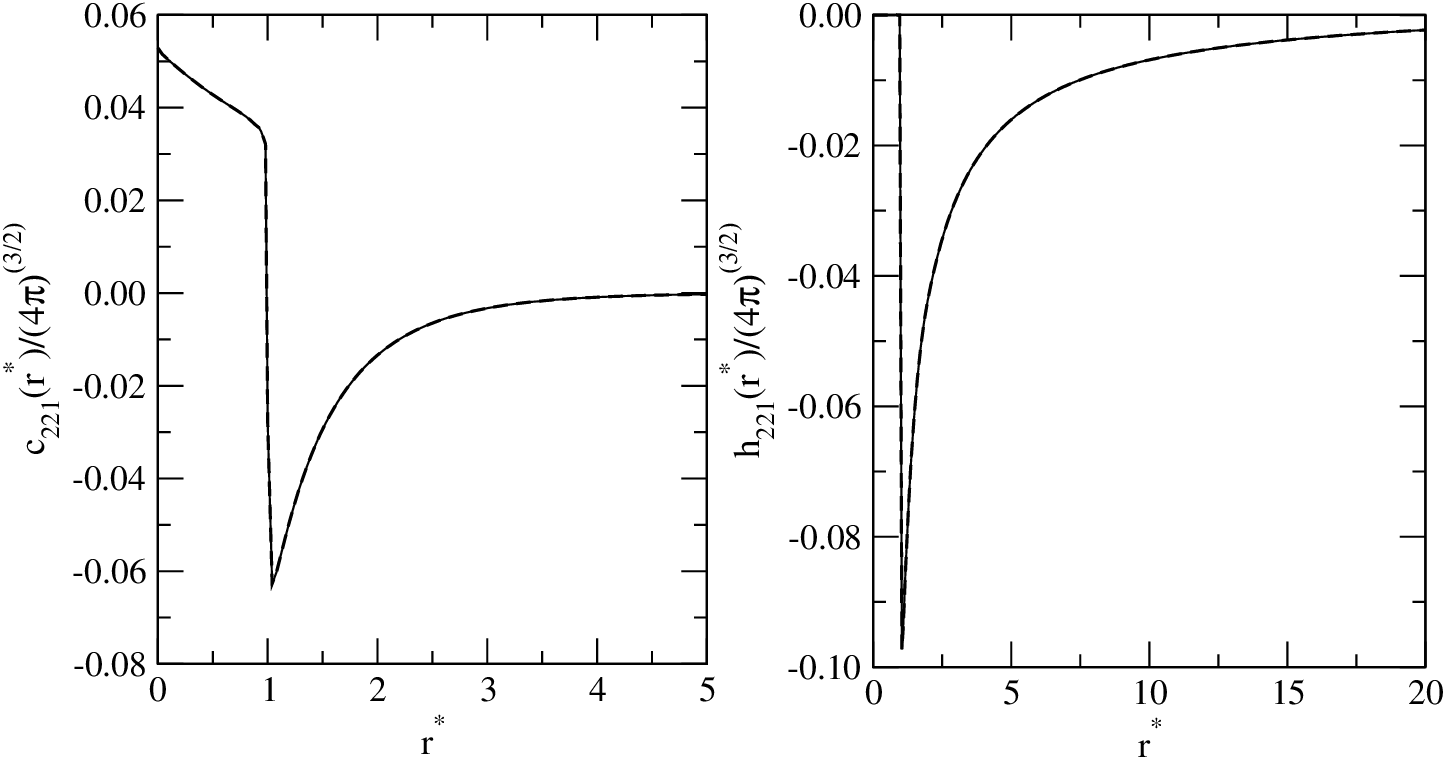}
\caption{Comparison of the harmonic coefficients $c_{221}(r^*)$ and $h_{221}(r^*)$ obtained by the analytical
solution[14](dashed line) with the results found using analytical method (full line) for the MSA. Parameters are same as
in Figure 1. A $1/r^*$ tail in harmonic coefficient $h_{221}(r^*)$ is seen.}
\end{center}
\end{figure}
                                                                                                                             
\begin{figure}[h]
\includegraphics[width=3.5in]{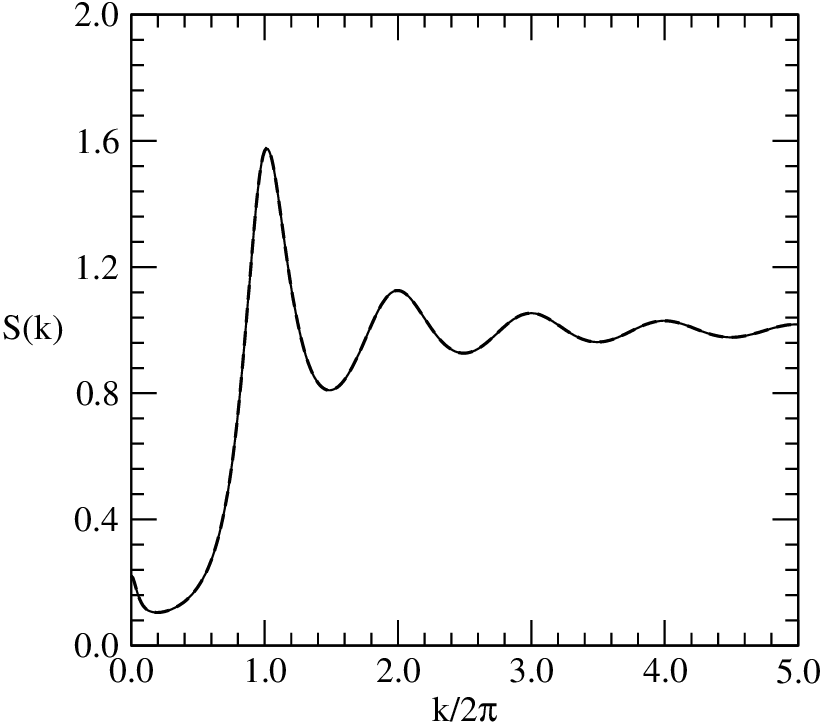}
\caption{ Comparison of the structure factor curves for the nematic phase
($\beta a_2=1, \beta a_0=0.1, z_0\sigma=z_2\sigma=1$, $\eta=0.315, P_2=0.63, P_4=0.27$). The dashed line is obtained with
the analytical solution[14] while full line is obtained by our numerical method for the MSA. The two curves overlap at
all values of $k$.}
\end{figure}

\begin{figure}[h]
\includegraphics[height=3.5in,width=5.0in]{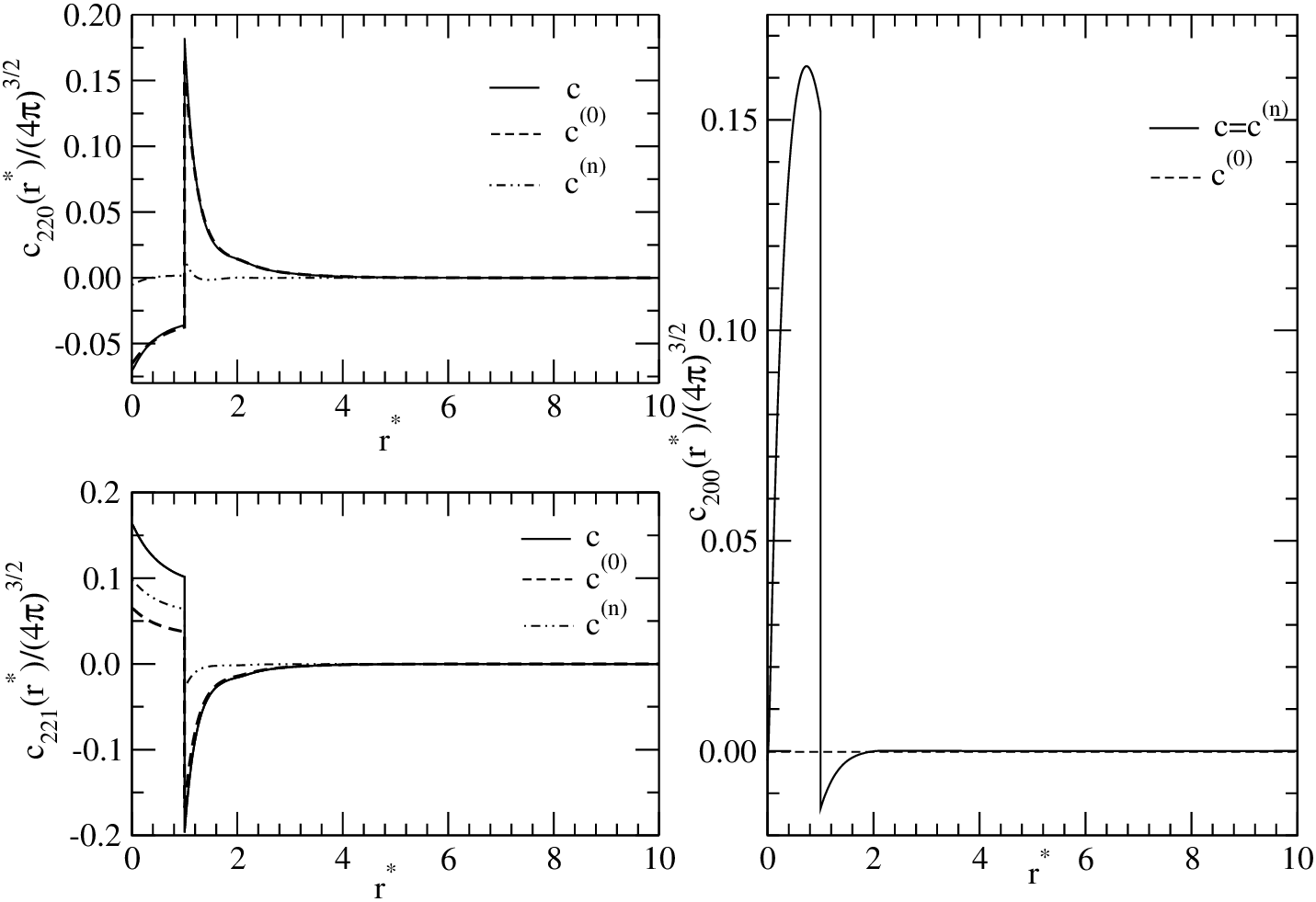}
\caption{Plot of the harmonic coefficients of the direct pair correlation function $c_{220}(r^*)$, $c_{221}(r^*)$
and $c_{200}(r^*)$ obtained by solving PY integral equation theory
($\beta a_2=1, \beta a_0=0.1, z_0\sigma=z_2\sigma=1$, $\eta=0.30$). While the contribution of symmetry breaking
part in $c_{221}(r^*)$ is small (shown by dot-dashed line), in $c_{221}(r^*)$ it is comparable inside the core.
The harmonic coefficient $c_{220}(r^*)$ arises due to symmetry breaking only.}
\end{figure}

\subsection{MSA Closure}

The MSA relation is written as 
\ba
h({\bf x}_1, {\bf x}_2)& = &-1,{\hspace{1cm}} |{\bf r}_2-{\bf r}_1|<{\sigma} \\
c({\bf x}_1, {\bf x}_2)& = & -\beta v(r, \Omega_1, \Omega_2),  {\hspace{0.5cm}} |{\bf r}_2-{\bf r}_1|>{\sigma}
\ea
where $v$ is given by Eqs.(2.4) and (2.5).

Condition (2.14) is exact for potential model of Eq(2.1) since $g(\br,\bom_1, \bom_2)=0$ for $r<\sigma$.
However, it is only for large $r$ that $c({\br,\bom_1, \bom_2})$ is asymptotic to $\beta v(r, \Omega_1, \Omega_2)$
but in MSA it is assumed that $c=-\beta v$ for all $r(r>\sigma)$.

Using Eq.(2.7) with the condition $l=m=0$ to expand the pair correlation functions and Eq(2.14) we get for 
$r<\sigma$

\ba
c_{000}(r)&=&-\gamma_{000}(r)-{(4\pi)^{3/2}}
\ea
and
\ba
c_{l_1l_2m}{(r)}=-\gamma_{l_1l_2m}{(r)}
\ea
For $r>\sigma$ from Eqs.(2.15), (2.4) and (2.5) we get
\be
c_{iim}(r)=(-1)^{m}\frac{(4\pi)^{3/2}}{2i+1}\frac{\beta a_i(z_i\sigma)^2\exp(-z_i r_{12})}{r/\sigma}
\ee
where $i=0,2$ and $m=0,1,2$.

\subsection{The PY Closure}

The PY relation is written as 
\be\
c({\bf x}_1, {\bf x}_2) = f({\bf x_1},{\bf x_2})[g({\bf x}_1, {\bf x}_2)-c({\bf x}_1, {\bf x}_2)]
\ee\
where $f({\bf x}_1, {\bf x}_2)=\exp[-\beta v({\bf x}_1, {\bf x}_2)]-1$
is the Mayer function and $\beta=(k_B T)^{-1}$; $k_B$ being the Boltzmann constant and T, temperature.
Expansion in spherical harmonics with constraint $l=m=0$ leads to 
\ba
c_{l_1l_2m}(r)={\frac{1}{\sqrt{4\pi}}}\sum_{l_{1}^{'}l_{2}^{'}m^{'}l_{1}^{''}l_{2}^{''}m^{''}}
f_{l_{1}^{'}l_{2}^{'}m^{'}}(r)&&[\gamma_{l_{1}^{''}l_{2}^{''}m^{''}}(r)+(4\pi)^{3/2}\delta_{l_{1}^{''},0}
\delta_{l_{2}^{''},0}\delta_{m^{''},0}]\nonumber \\
&&\Gamma_{m^{'}m^{''}m}^{l_{1}^{'}l_{1}^{''}l_{1}}{\hspace{0.3cm}}\Gamma_{{\underline m}^{'}{\underline m}^{''}
{\underline m}}^{l_{2}^{'}l_{2}^{''}l_{2}}
\ea
where $f_{l_{1}^{'}l_{2}^{'}l^{'}}(r)$ is the harmonic coefficient of the Mayor function
$f(r,\Omega_1, \Omega_2)$.
 
We solved the OZ equation for both the MSA and the PY closures for given values of order parameters $P_2$ and
$P_4$. In order to solve these equations numerically we followed the iterative method described in ref.[13].
However, as coefficients $h^{(n)}_{l_1l_2m}(r)$ may decay slowly and extend
to relatively large values of $r^*(=\frac{r}{\sigma})$ in nematics we have extended the range of $r^*$(i.e. $r^*=60$)
to ensure proper convergence. The other point which needed special care is related to the pronounced long range
tail which occurs in coefficients $h^{(n)}_{l_1l_2m}(r^*)$ with $m=\pm 1$ (see Fig 6). Before performing the
Hankel transform in each iteration we fit the data points of these harmonic coefficients beyond $r^*>r^{*}_{0}$(=
20) to a power law $a+\frac{b}{r^*}$, shift them by $a$ and then extrapolate them to 
infinity [19]. This removes the finite size effect on the tail. 

The potential parameters taken in our calculations are $z_0\sigma=z_2\sigma=1$, $\beta a_0=0.1$ and $\beta a_2=1$.
For the PY we have also considered the case of $\beta a_2=0.5$.

In Fig.1 we compare the results of the Fourier transform of some of the harmonic coefficients $h_{l_1l_2m}(r^*)$
obtained with the analytical solution of the model potential [14] with the results found using numerical
method stated above for the MSA. Both results are for $\eta(= \frac{\pi\sigma^*}{6})=0.315, 
P_2=0.63$ and $P_4=0.27$. The 
minimization of free energy functional (see Sec III) which contains both the symmetry breaking and
symmetry conserving parts of the direct pair correlation function gives these values of order parameters $P_2$ and
$P_4$. These values of order parameters are also found from the analytical result of ref[14].
Both curves shown in the figure overlap indicating an excellent agreement between the two results. In Fig.2 we compare the harmonic 
coefficients $c_{221}(r^*)$ and $h_{221}(r^*)$. These harmonic coefficients are 
of fundamental importance as they appear in nematic elastic constants. The decay of $h_{221}(r^*)$
as $\frac{1}{r^*}$ at large distance is clearly seen. In Fig.3 we compare our results of the structure factor 
defined as 
\ba
S(k)&=&1+ \rho^{*} \int f(\Omega_1)h(k, \Omega_1, \Omega_2) f(\Omega_2) d{\Omega_1}d{\Omega_2}\nonumber \\
&=& 1+ {\frac{\rho^{*}}{(4\pi)^{3/2}}}[h_{000}(k)+2\sqrt{5}P_2h_{200}(k)+5P_{2}^{2}h_{220}(k)].
\ea

Again we find excellent agreement between analytical and numerical results including small peak at k=0 which is 
attributed to the appearance of additional effective attraction due to parallel alignment of molecules [14]. 

In Figs 4-6 we give results found from using the PY closure for $\beta a_2=1, \eta=0.30, P_2=0.69$
and $P_4=0.32$ in the director space. These values of order parameters have been found 
from the minimization of the free energy functional (see Sec III). While the harmonic coefficients
$c_{220}(r^*)$ and $c_{221}(r^*)$ shown in Fig 4 survive both in the isotropic
$P_2=P_4=0$ and in the nematic phase $P_2\neq 0, P_4\neq 0$, the harmonic coefficient $c_{200}(r^*)$
survive only in the nematic phase and vanishes in the isotropic phase. The contribution arising due to 
symmetry breaking to the harmonic coefficients $c_{220}(r^*)$ and $c_{221}(r^*)$ shown in Fig 4 by 
dot-dashed line are found to be very small 
compared to the symmetry conserving part. Few selected harmonic coefficients of $h$ are shown in Figs 5 and 6 in the
director space. In Fig 5 we plot the harmonic coefficients $h_{200}(r^*)$ and $h_{220}(r^*)$. While
$h_{200}(r^*)$ survive only in the nematic phase, $h_{220}(r^*)$ survive both in the isotropic and in 
the nematic. In the case of $h_{220}(r^*)$ we also plot the contributions arising  due to symmetry breaking and symmetry 
conserving and note that the contribution arising due to symmetry breaking is small. 
In Fig 6 we plot harmonic coefficients $h_{221}(r^*)$ and show its $\frac{1}{r^*}$ dependence in the inset.

In Figs 7 and 8 we plot few selected harmonic coefficients of $c$ and $h$ in 
director space for $\beta a_2=0.5$, $\eta=0.48$, $P_2=0.54$ and $P_4=0.12$. As will be shown later that at the 
isotropic-nematic transition the packing fraction $\eta$ of the nematic phase is 0.458. for $\beta a_2=0.5$ while at 
$\beta a_2=1$ it is 0.244. The comparison of these harmonic coefficients show that 
orientational ordering has more pronounced 
effect on these harmonic coefficients when $\beta a_2=1$ compared to that of $\beta a_2=0.5$. 
This could be easily understood from the fact that the orientational ordering arises solely due to the
long range anisotropic part of the interaction and $\beta a_2$ measures its strength.
We note that the PY closure gives harmonic coefficients of both symmetry breaking and 
symmetry conserving parts of pair correlation functions which have features similar 
to those found from the MSA solution as well as from computer simulation [13].
\begin{figure}[h]
\begin{center}
\vspace{1.3cm}
\includegraphics[height=2.8in,width=5.0in]{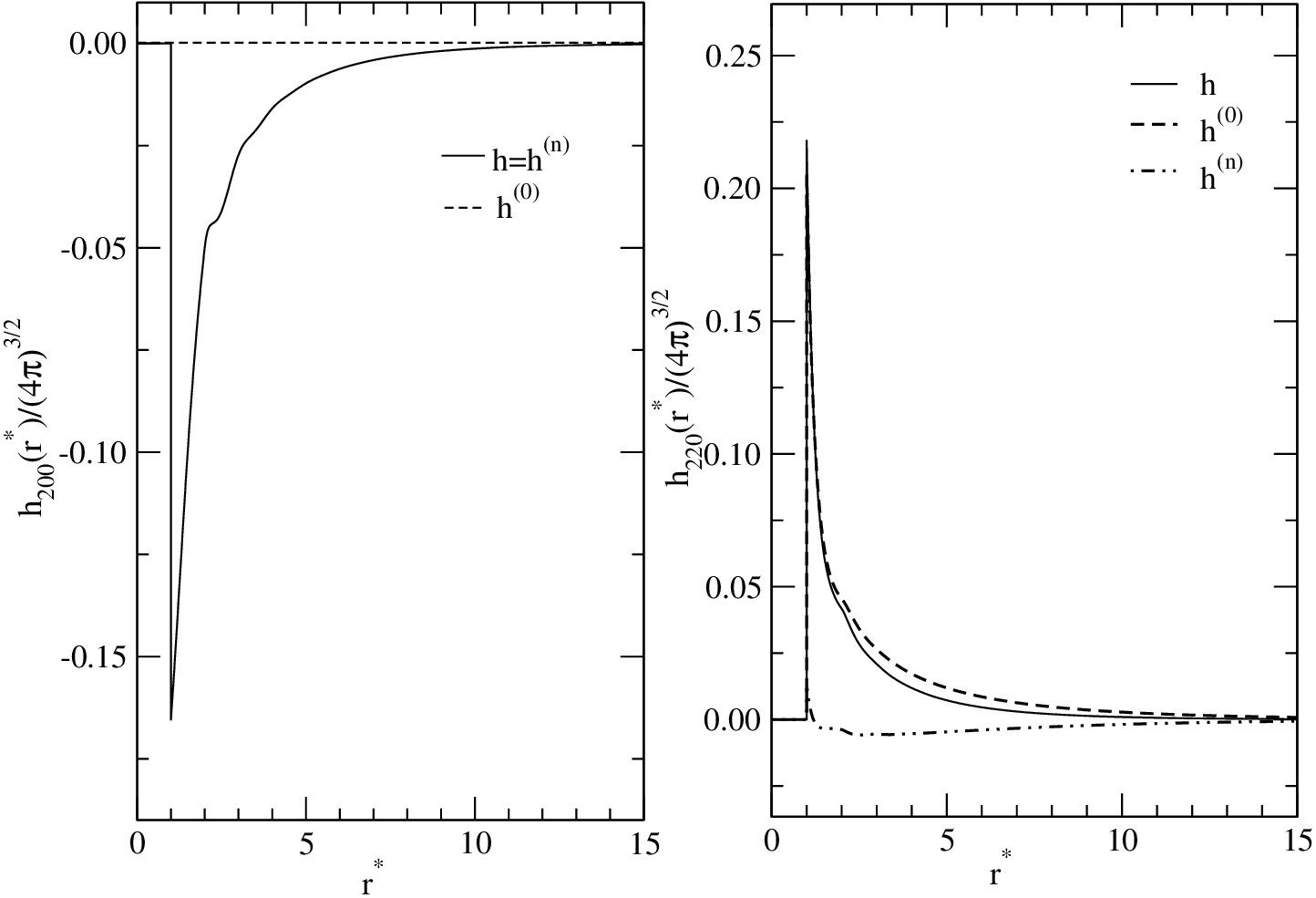}
\caption{Coefficients $h_{200}(r^*)$ and $h_{220}(r^*)$ obtained from the PY theory with the parameters same as
in Figure 4. The symmetry breaking contribution shown by dot-dashed line is very small for $h_{220}(r^*)$.
The harmonic coefficient $h_{200}(r^*)$ arises due to symmetry breaking only.}
\end{center}
\end{figure}
                                                                                                                             
\begin{figure}[h]
\begin{center}
\includegraphics[height=3.5in,width=4.0in]{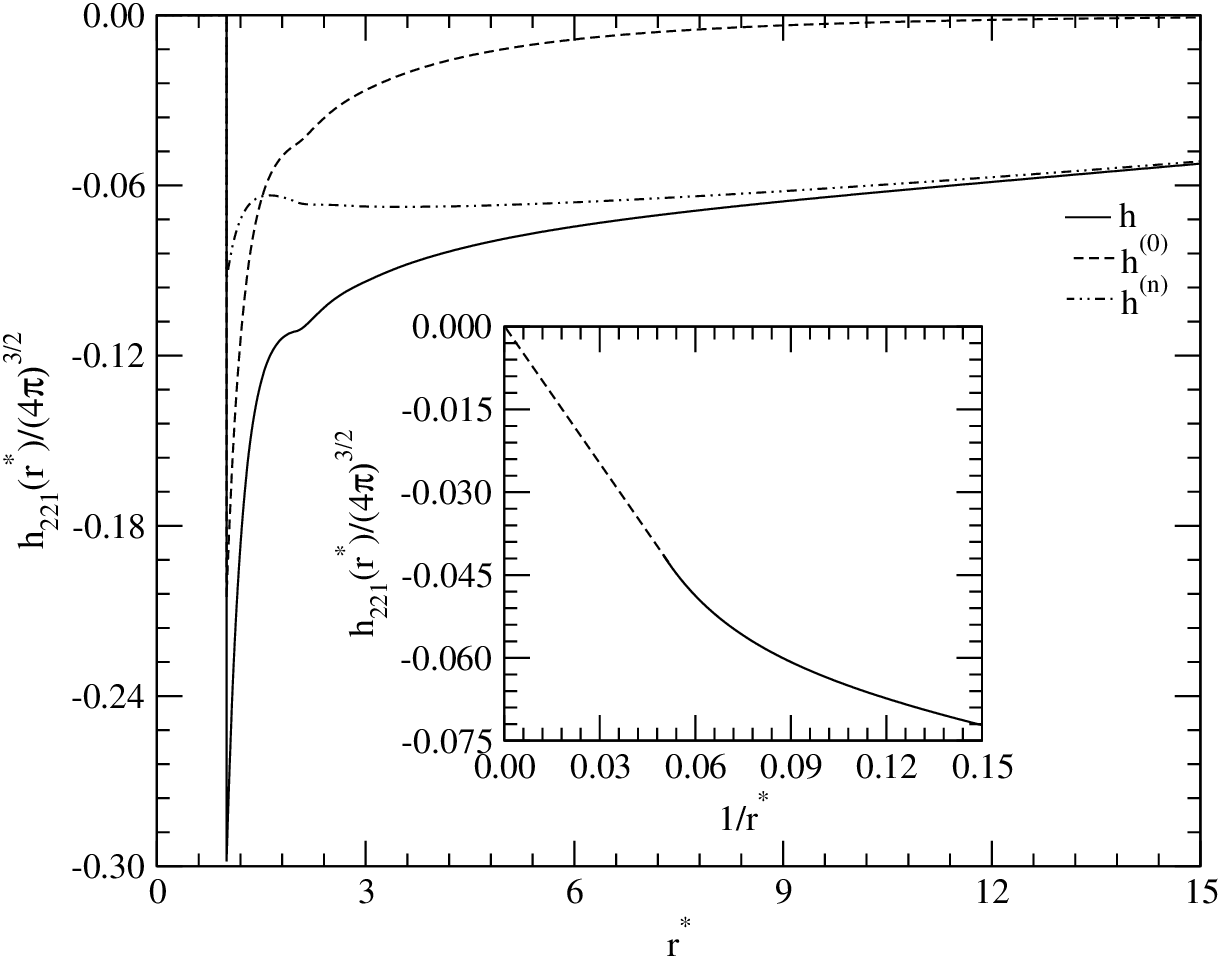}
\caption{Harmonic coefficient $h_{221}(r^*)$ in the director frame. Details are same as in Figure 4. Inset shows the plot
of $h_{221}(r^*)$ with respect to $1/r^*$; the dashes line shows the extrapolated part. The origin of tail
 is due to orientational symmetry breaking.}
\end{center}
\end{figure}
                                                                                                                             
\begin{figure}[h]
\begin{center}
\includegraphics[height=3.5in,width=6.0in]{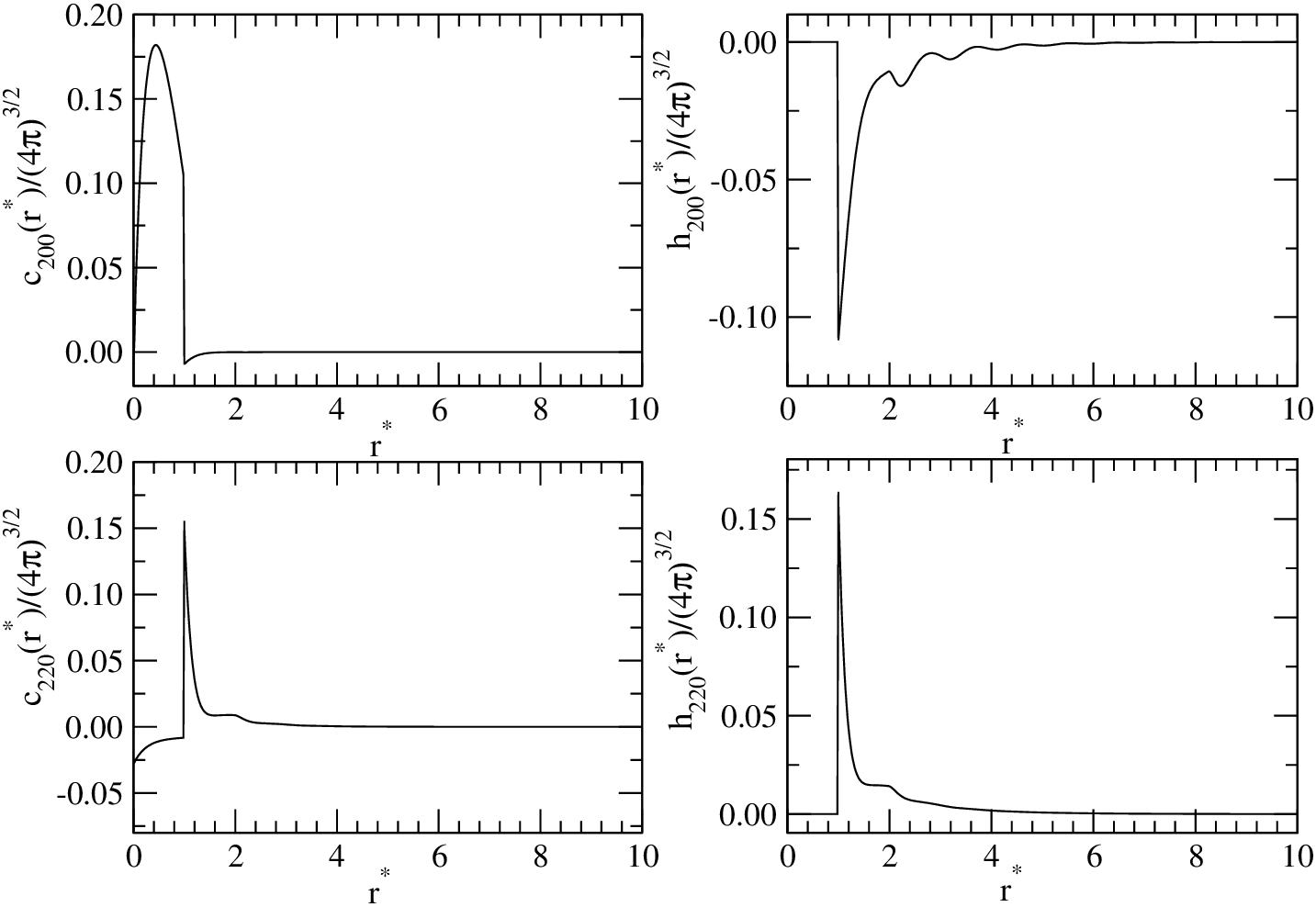}
\caption{Plot of the symmetry breaking harmonic coefficients ($c_{200}(r^*)$, $h_{200}(r^*)$) and symmetry
conserving harmonic coefficients ($c_{220}(r^*)$, $h_{220}(r^*)$) found by PY integral equation theory with
the parameters $\beta a_2=0.5, \beta a_0=0.1, z_0\sigma=z_2\sigma=1$, $\eta=0.48$.}
\end{center}
\end{figure}

\begin{figure}[h]
\begin{center}
\includegraphics[height=3.5in,width=5.0in]{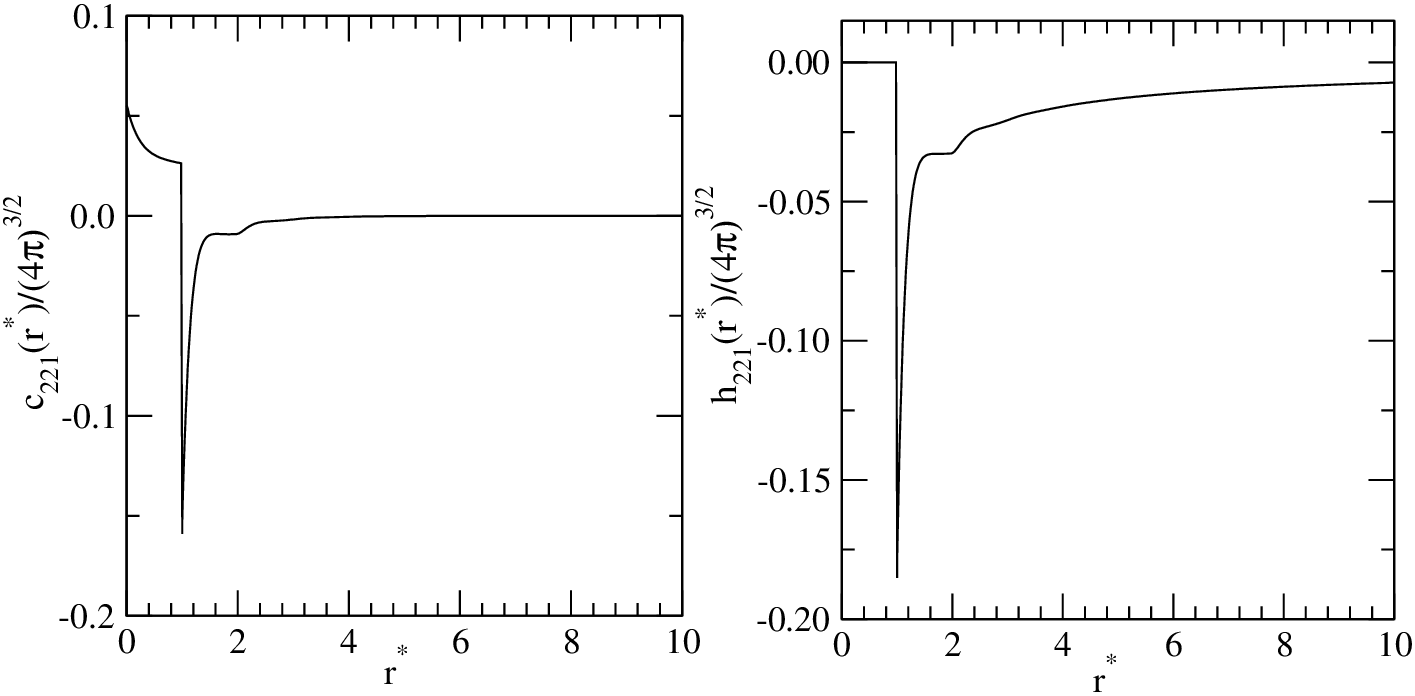}
\caption{ Harmonic coefficients $c_{221}(r^*)$ and $h_{221}(r^*)$ obtained by PY theory with the parameters same
as in Figure 7.}
\end{center}
\end{figure}

In Fig 9 we plot the structure factor found from the PY theory. In this case its expression
is found to be 
\ba
S(k)&=& 1+ {\frac{\rho}{(4\pi)^{3/2}}}\sum_{l_1l_2}\sqrt{(2l_1+1)(2l_2+1)}P_{l_1}P_{l_2}h_{l_1l_20}(k)
\ea
The curve drawn in full line corresponds to $\eta=0.30, P_2=0.69, P_4=0.32, \beta a_2=1$ whereas the one drawn in dashed line corresponds to $\eta=0.48, P_2=0.54, P_4=0.12$ at $\beta a_2=0.5$. We note that 
$\eta=0.48$ is close to the freezing transition where the system goes into the crystalline phase 
and therefore the peaks in $S(k)$ are more pronounced compared to the curve corresponding to $\eta=0.30$.
According to Hansen-Verlet[25] criterion fluid becomes unstable when the height of 
main peak in $S(k)$ becomes equal to $2.9\pm 0.1$. The curves corresponding to $\eta=0.30$, $\beta a_2=1$
shows a small peak at $k=0$ as was found in the case of the MSA theory. However, this peak is not seen 
in the curve corresponding to $\eta=0.48, \beta a_2=0.5$. For this case as indicated by the values of order parameters, the orientational ordering is weak compared to that of $\eta=0.30, \beta a_2=1$ and therefore the 
effective attraction which arises due to orientational ordering is negligible. 

\section{Free energy functional}
The reduced free energy $A[\rho]$ of an inhomogeneous system is a functional of $\rho({\bf x})$
and is written as[21]
\be
A[\rho]=A_{id}[\rho]+A_{ex}[\rho]
\ee
The ideal gas part is exactly known and is given as 
\be
A_{id}[\rho]=\int d{\bf x}\rho({\bf x})[ln(\rho({\bf x})\Lambda)-1]
\ee
where $\Lambda$ is cube of the thermal wavelength associated with a molecule. 
The excess part arising due to intermolecular interactions is related to 
the direct pair correlation function(DPCF) as 
\ba
{} \hspace {-0.15in} \frac{\delta^2 A_{ex}}{\delta \rho({\bf x_1})\delta\rho({\bf x_2})}=-c^{(0)}({\bf{x_1, x_2}}; \rho_0)-c^{(n)}(\bf{x_1, x_2}; [\rho])
\ea
where superscripts $(0)$ and $(n)$ represent respectively the symmetry conserving and symmetry breaking 
parts of the DPCF. In other words, $c^{(0)}$ is found by putting 
order parameters in Eqs(2.13) equal to zero whereas $c^{(n)}$ are the contributions which arise 
when order parameters are nonzero.

$A_{ex}$ is found by functional integration of Eq.(3.3). In this integration the system
is taken from some initial density to the final density $\rho(x)$ along 
a path in the density space; the result is independent of the path of integration[26].
For the symmetry conserving part $c^{(0)}$ the integration in density space is
done taking isotropic fluid of density $\rho_l$ (the density of coexisting fluid)
as reference. This leads to
\ba
A_{ex}^{(0)}[\rho]&=&A_{ex}(\rho_l)-\frac{1}{2}\int d{\bf x_1}\int d{\bf x_2}
\Delta\rho(\bf {x_1})\Delta\rho({\bf x_2}) \nonumber \\
&& \times {\bar c^{(0)}}({\bf x_1,x_2})
\ea
where
\ba
{\bar c^{(0)}}({\bf x_1,x_2})&=& 2\int d\lambda \lambda \int
d\lambda^{'} c^{(0)}\{{\bf x_1,x_2}; \rho_l+\lambda\lambda^{'}(\rho_0-\rho_l)\} \nonumber \\
{\rm and}\nonumber \\
 \Delta\rho({\bf{x}}) &=&\rho({\bf{x}})-\rho_l 
\ea
 $A_{ex}(\rho_l)$ is the excess reduced free energy of isotropic fluid of density
$\rho_l$ and $\rho_0$ is the average density of the ordered phase.

In order to integrate over $c^{(n)}[\rho]$, we characterize the density
space by two parameters $\lambda$ and $\xi$ which vary from 
$0$ to 1[19]. The parameter $\lambda$ raises density from $0$ to $\rho_0$ as
it varies from 0 to 1 whereas parameter $\xi$ raises the order parameter from 
$0$ to $P_l$ as it varies from 0 to 1. If we have $n$ order parameters to describe the 
ordered phase we can think of a $n-$dimensional order parameter space; a point in this space
defines the values of the $n$ order parameters.
The integration over $c^{(n)}(\rho)$ in Eq(3.3) can be done along a straight
line path that connects origin to a point corresponding to the final values of all $n$ order parameters.
This path is characterized by the variable $\xi$. This gives 
\be
A_{ex}^{(n)}[\rho]=-\frac{1}{2}\int d{\bf x_1} \int d{\bf x_2} \rho({\bf x_1}) \rho({\bf x_2})
{\tilde c^{(n)}}({\bf x_1,x_2})
\ee
where
\ba
{\bar c^{(n)}}({\bf x_1,x_2}) &=&  4\int_{0}^{1} d\xi \xi \int_{0}^{1} d\xi^{'}
\int_{0}^{1} d\lambda \lambda \int_{0}^{1}d\lambda^{'} \times \nonumber \\
& & c^{(n)}({\bf x_1, x_2}, \lambda\lambda^{'}\rho_0; \xi\xi^{'}\sqrt{\sum_{l=1}^{n}}{P_l}^{2}).
\ea

While integrating over $\lambda$ the order parameters $P_l$ are 
kept fixed and while integrating over $\xi$ the density is 
kept fixed. The result does not depend on the order of integration.
The free energy functional of an ordered phase is the sum of  $A_{id}$, $A_{ex}^{(0)}$ and $A_{ex}^{(n)}$
given respectively by Eqs(3.2), (3.4) and (3.6). Note that the Ramakrishnan
and Youssouff [23] free energy functional is the sum of only $A_{id}$ and $A_{ex}^{(0)}$ and
contains an additional approximation in which ${\bar c}^{(0)}({\bf x_1,x_2})$ in (3.4) is replaced by
$c({\bf x_1,x_2}; \rho_l)$.

The minimization of $\Delta A= A[\rho]-A(\rho_{0})$ where $A(\rho_{0})$ is the free energy of an isotropic phase of density
$\rho_{0}$ leads to 
\be
\ln f(\Omega)=C+\int d{\bf x}_{2} \Delta \rho({\bf x}_{2}){\tilde c}^{(0)}({\bf x}_{1},{\bf x}_{2},
\rho_{0})+\int d{\bf x}_{2} \rho({\bf x}_{2}){{\tilde c}_{1}^{(n)}({\bf x}_{1},{\bf x}_{2})}
\ee
where $f(\Omega)=\frac{\rho(x)}{\rho_{0}}$
and
\be
{\tilde c_{1}^{n}}({\bf x}_{1},{\bf x}_{2})=2\int_{0}^{1} d\lambda \lambda \int_{0}^{1} d\lambda^{'} 
\int_{0}^{1} d\xi c^{(n)}({\bf x}_{1},{\bf x}_{2}; \lambda \lambda^{'}\rho_0;\xi\sqrt{\sum_{l=1}^n}{P_l}^{2})
\ee
\ba
{{\tilde c}^{(0)}}({\bf x_1,x_2})&=& \int d\lambda 
c^{(0)}\{{\bf x_1,x_2}; \rho_l+\lambda(\rho_0-\rho_l)\} \nonumber \\
\ea
The constant C is found from the normalization condition
\be
\int f(\Omega)d\Omega=1
\ee
In order to evaluate ${\bar c}^{(n)}({\bf x}_1, {\bf x}_2)$ and  ${\tilde c}_{1}^{(n)}(\bf{x}_1, \bf {x}_2)$ 
from Eqs.(3.7) and (3.9) we 
need symmetry breaking part of DPCF from density 
zero to $\rho_0$ and order parameters form zero to $P_l$ at sufficiently 
small intervals. The computational time needed to evaluate these correlation functions depends on
the number of order parameters one takes in the calculation. A nematic is adequately 
described by two order parameters, $P_2$ and $P_4$. However, ordered phases such as
smectic and crystalline solids may need several order parameters. It is therefore
advisable to approximate the values of $A^{(n)}_{ex}[\rho]$ with as small number 
of order parameters as possible. Here we show that for nematic it is a good approximation 
to consider only $P_2$ in calculating $A^{(n)}_{ex}[\rho]$ from Eq.(3.6).

In case of the MSA the free energy functional reduces to
\ba
\frac {A[\rho(x)]-A[\rho_0]}{N}=\frac{\Delta A[\rho]}{N} & = & -\ln Z+ [P_{2}^{2}
{\hat{\tilde c}}_{220}^{(0)}(0)+P_{2}{\hat{\tilde c_{1}}}_{200}^{(n)}(0)+P_{2}^{2}
{\hat{\tilde c_{1}}}_{220}^{(n)}(0)]\nonumber \\
& & -\frac{1}{2}P_{2}^{2}{\hat{\bar c}}_{220}^{(0)}(0)-\frac{1}{2}[{\hat{\bar c}}_{000}^{(n)}(0)+P_{2}
({\hat{\bar c}}_{200}^{(n)}(0)+{\hat{\bar c}}_{020}^{(n)}(0))\nonumber \\
& &+P_{2}^{2}{\hat{\bar c}}_{220}^{(n)}(0)].
\ea
where
\be
Z=\frac{1}{2}\int_{-1}^{1}d\cos(\theta) \exp[(P_2{\hat{\tilde c}}_{220}^{(0)}(0)+{\hat{\tilde c_{1}}}_{200}^{(n)}(0)
+P_{2}
{\hat{\tilde c_{1}}}_{220}^{(n)}(0))P_{2}(\cos \theta)]
\ee
and
\ba
{\hat {\bar c}}^{(i)}_{l_1l_20}(0)&=&\frac{\rho_0}{\sqrt{4\pi}}{\sqrt{(2l_1+1)(2l_2+1)}}\int_{0}^{\infty} 
{\bar c}^{(i)}_{l_1l_20}(r)r^2 dr. \nonumber 
\ea
Note that in this case $P_4$ does not appear explicitly. It appears only through ${\bar c}^{(n)}$
and ${\tilde c}_{1}^{(n)}$. We calculated ${\bar c}^{(n)}$ and ${\tilde c}_{1}^{(n)}$ 
at $\eta=0.315$ using the values of the DPCF obtained for 
$\eta$ from zero to 0.315 at the interval of 0.02, $P_2$ from zero to 0.70 and $P_4$ from zero
to 0.35 at the interval of 0.05. Substituting the values of 
${\bar c}^{(0)}$(which correspond to $P_2=P_4=0$) , ${\bar c}^{(n)}$
and ${\tilde c}_{1}^{(n)}$ in Eq. (3.12) we minimized the free energy with respect to 
$P_2$. The order parameter $P_4$ is found from the relation
\be
P_{4}=\frac{1}{2Z}\int d\cos(\theta){\exp[(P_2{\hat{\tilde c}}_{220}^{(0)}(0)+
{\hat{\tilde c_{1}}}_{200}^{(n)}(0)+P_{2}{\hat{\tilde c_{1}}}_{220}^{(n)}(0))
P_{2}(\cos \theta)]P_4(\cos \theta)},
\ee
The values found are $P_2=0.63$, $P_4=0.27$. The values of structural 
parameters defined as
\be
{\hat c}_{l_1l_20}(0)=\frac{\rho_0}{\sqrt{4\pi}}{\sqrt{(2l_1+1)(2l_2+1)}}\int_{0}^{\infty} 
c_{l_1l_20}(r)r^2 dr \nonumber 
\ee
and
\be
{\hat c}_{l_1l_21}(0)={\rho_0}\int_{0}^{\infty} c_{l_1l_21}(r)r^2 dr \nonumber 
\ee
are ${\hat c}_{220}(0)=5.29$ and ${\hat c}_{221}(0)=-3.60$. 

We next calculate ${\bar c}^{(n)}$ 
and ${\tilde c}_{1}^{(n)}$ in same way except taking $P_4=0$. When these values of ${\bar c}^{(n)}$ 
and ${\tilde c}_{1}^{(n)}$ were used we found
$P_2=0.65, P_4=0.28, {\hat c}_{220}(0)=5.46$ and ${\hat c}_{221}(0)=-3.62$. The two
set of values compare well and indicate that using only principal order parameter
in calculating $A_{ex}^{(n)}$ is a good approximation.

The Ward identity which must be satisfied in a nematic phase relates the single particle 
distribution to an integral of direct pair correlation function. When this identity is 
expressed in a functional differential form it reduces to the Lovett equation[17]
\be
{\nabla_{\Omega_1}}{\ln\rho(x_1)}=\int c(r,\Omega_1,\Omega_2) {\nabla_{\Omega_2}} \rho(x_2)d\br d\bom_2
\ee
Expanding it in spherical harmonics we get 
\be
1=-\rho{\sum_{l_1l^{'}}}\sqrt{\frac{(2l_1+1)}{20\pi}}{(2l^{'}+1)} P_{l^{'}}C_g(l_1l^{'}2000)
C_g(l_1l^{'}2101)\int dr r^{2}c_{l_121}(r)
\ee
When the two sets of parameters reported above, one in which both $P_2$ and $P_4$ appeared in calculating
the pair correlation functions while in other only $P_2$ appeared, are substituted in this equation 
we find that it is satisfied with accuracy better than $10^{-3}$. From these results
we conclude that it is sufficient to evaluate $\bar c^{(n)}$ and ${\tilde c}_{1}^{(n)}$
with principal order parameter only. All results given below correspond to this approximation.

For the PY the free energy functional is found to be 
\ba
\frac {A[\rho(x)]-A[\rho_0]}{N}=\frac{\Delta A[\rho]}{N} & = & -\ln Z+ [P_{2}^{2}
{\hat{\tilde c}}_{220}^{0}(0)+P_{2}{\hat{\tilde c_{1}}}_{200}^{(n)}(0)+P_{2}^{2}{\hat{\tilde c_1}}_{220}^{(n)}(0)+P_2P_4
{\hat{\tilde c_{1}}}_{240}^{(n)}(0) \nonumber \\
& &+P_{4}^2{\hat{\tilde c}}_{440}^{(0)}(0)+P_4{\hat{\tilde c_{1}}}_{400}^{(n)}(0)
+P_2 P_4{\hat{\tilde c_{1}}}_{420}^{(n)}(0)+P_{4}^{2}
{\hat{\tilde c_{1}}}_{440}^{(n)}(0)]\nonumber \\
& & -\frac{1}{2}(P_{2}^{2}{\hat{\bar c}}_{220}^{(0)}(0)+2P_2P_4{\hat{\bar c}}_{240}^{(0)}(0)+
P_{4}^{2}{\hat{\bar c}}_{440}^{(0)}) (0) \nonumber \\
& & -\frac{1}{2}[{\hat{\bar c}}_{000}^{(n)}(0)+P_{2}
({\hat{\bar c}}_{200}^{(n)}(0)+{\hat{\bar c}}_{020}^{(n)}(0))
+P_{2}^{2}{\hat{\bar c}}_{220}^{(n)}(0) \nonumber \\
& &+2P_{4}{\hat{\bar c}}_{400}^{(n)}(0)+2P_2P_4{\hat{\bar c}}_{240}^{(n)}(0)+P_{4}^2{\hat{\bar c}}_{440}^{(n)}(0)].
\ea
with
\ba
Z&=&\frac{1}{2}\int_{-1}^{1}d\cos(\theta) \exp[\{P_2{\hat {\tilde c}}_{220}^{(0)}(0)+
{\hat{\tilde c_{1}}}_{200}^{(n)}(0)+P_{2}
{\hat{\tilde c_{1}}}_{220}^{(n)}(0)+P_4{\hat{\tilde c_{1}}}_{240}^{(n)}(0)\}P_{2}(\cos \theta)+ \nonumber \\
&& \{P_4{\hat{\tilde c}}_{440}^{(0)}(0)+{\hat{\tilde c_{1}}}_{400}^{(n)}(0)+P_2{\hat{\tilde c_{1}}}_{420}^{(n)}(0)+P_{4}
{\hat{\tilde c_{1}}}_{440}^{(n)}(0)\}P_4(\cos \theta)]
\ea

In these equations, unlike the MSA (see Eq.(3.12)), both $P_2$ and $P_4$ appear. The values of 
${\hat{\bar c}}_{l_1l_2m}^{(n)}(0)$ and ${\hat{\tilde c_{1}}}_{l_1l_2m}^{(n)}(0)$ have been 
calculated at $\eta=0.30$ using the values of the harmonic coefficients of 
$c^{(n)}$ obtained for $\eta$ from zero to 0.30 at the interval of 0.02 and 
the $P_2$ from zero to 0.75 at the interval of 0.05. We substituted these 
values of ${\bar c}^{(0)}$ and ${\bar c}^{(n)}$ and 
${\tilde c}_{1}^{n}$ in Eq(3.19) and minimized the resulting expression with respect to 
$P_2$ and $P_4$. The values of these order parameters and the values of the structural 
parameters at $\eta=0.30$ and $\beta a_2=1$ are found to be:

$P_2=0.69$, $P_4=0.32$, ${\hat{c}}_{220}(0)=5.64$, ${\hat{c}}_{221}(0)=-3.84$,
${\hat{c}}_{421}(0)=0.045$.

When these values were used in Eq.(3.15) the Ward identity was found to be adequately satisfied.

When we chose $\beta a_2=0.50$ then as shown in the following section the nematic 
packing fraction at the  nematic-isotropic transition is found to be $\eta =0.458$. 
We therefore calculated the free energy at $\eta =0.48$ which is in the nematic region.
In this case we found

$P_2=0.54$, $P_4=0.12$, ${\hat{c}}_{220}(0)=4.58$, ${\hat{c}}_{221}(0)=-3.11$ ,${\hat{c}}_{421}(0)=0.048$. 

These values also satisfy the Ward identity.
\begin{figure}[h]
\begin{center}
\vspace{1.2cm}
\includegraphics[width=4.0in]{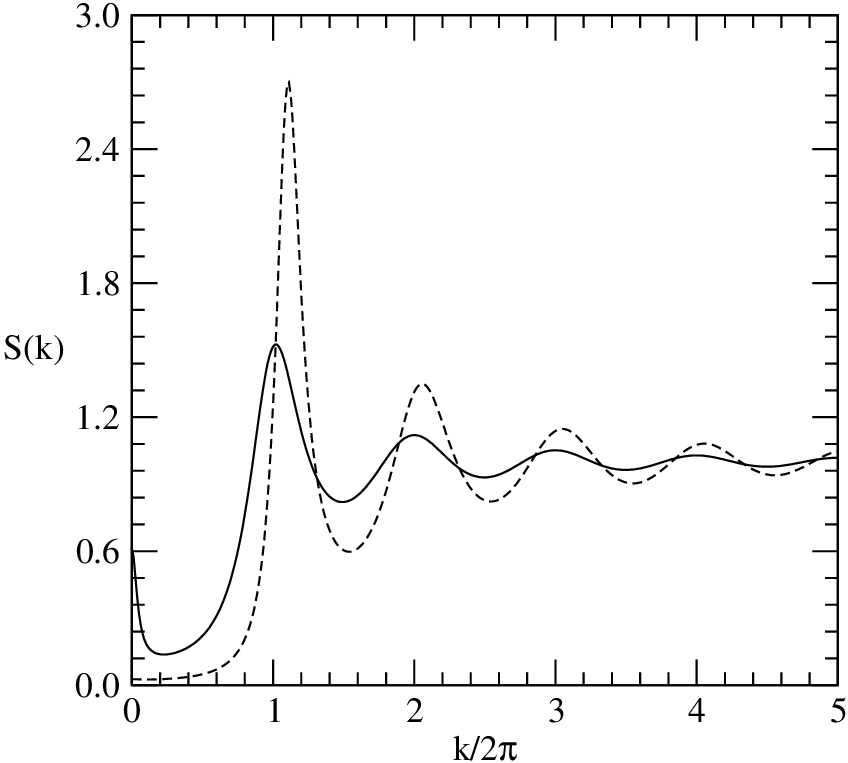}
\caption{Plots of structure factor obtained by using PY theory for
$\beta a_2=1$ ($\eta=0.30, P_2=0.69, P_4=0.32$) (full line) and $\beta a_2=0.5$
($\eta=0.48, P_2=0.54, P_4=0.12$) (dashed line). Other potential parameters are same as in figure 7.
A small peak at $k=0$ exists for $\beta a_2=1$ but not for $\beta a_2$=0.5.}
\end{center}
\end{figure}
                                                                                                                             
\begin{figure}[h]
\includegraphics[width=4.0in]{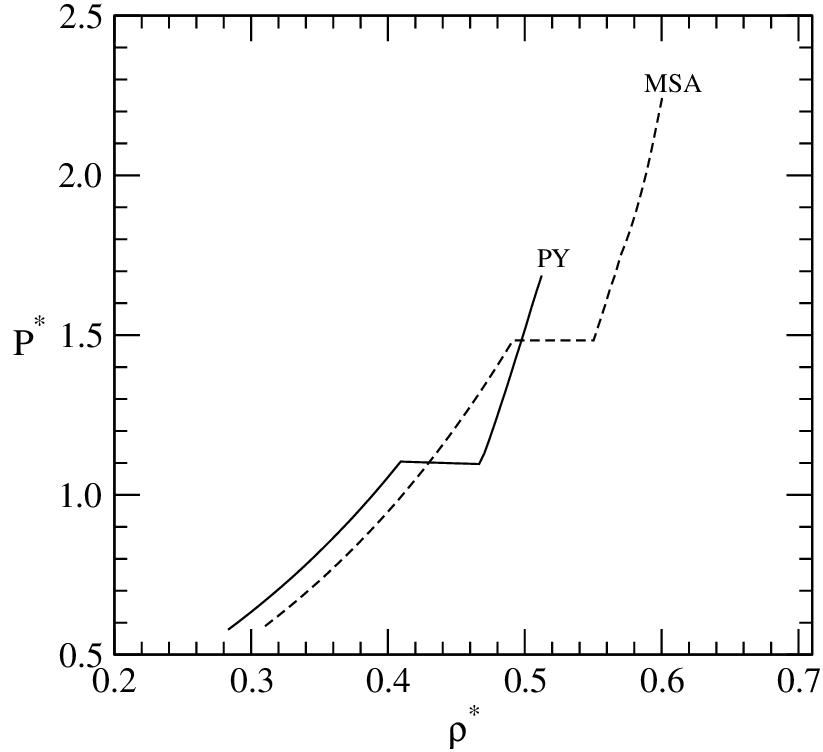}
\caption{$P^*-\rho^*$ isotherms for the MSA and PY closures for $\beta a_2=1$.
Other potential parameters are same as in Figure 1. The plateau corresponds to change in densiy at the
isotropic-nematic transition. }
\end{figure}
                                                                                                                             
\begin{figure}[h]
\includegraphics[width=4.0in]{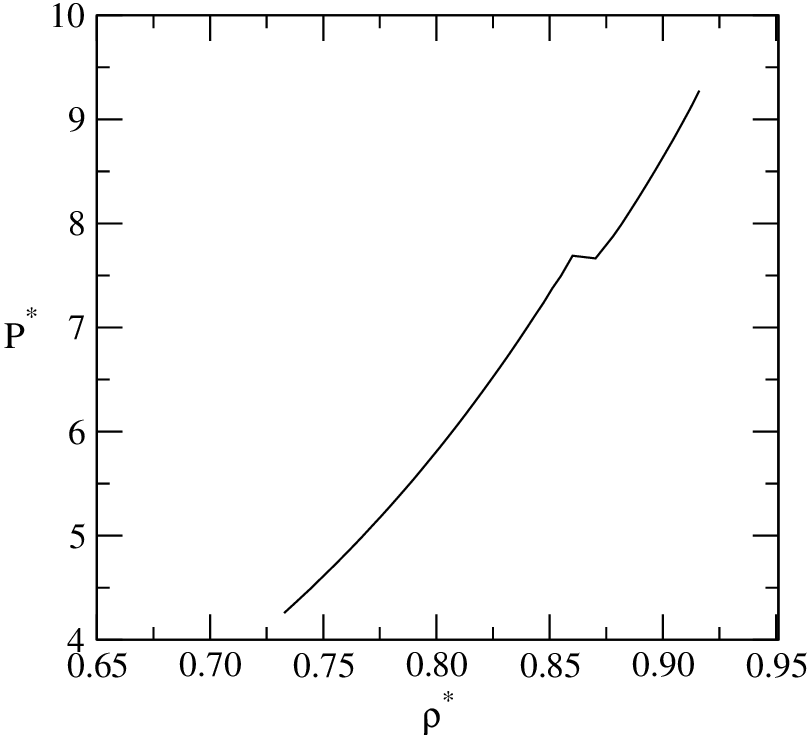}
\caption{$P^*-\rho^*$ isotherm obtained for the PY closure with potential parameter $\beta a_2=0.5$. Other
potential parameters are same as in figure 1.}
\end{figure}
\section{Isotropic-nematic transition}
The grand thermodynamic potential, defined as 
\be
W=A-\beta \mu \int d{\bf x} \rho({\bf x})
\ee
where $\mu$ is the chemical potential, is generally preferred to locate 
the freezing transition as it ensures that the pressure and chemical potential 
of the two phases at the transition remains equal. Using Eqs.(3.1)-(3.6) and (4.1) we get
\ba
-\Delta W=W_l-W_f&=& \int d{\bf x_1}[\rho({\bf x_1})\ln{\frac{\rho(x_1)}{\rho_f}}-\Delta\rho(x_1)]\nonumber \\
& & -\frac{1}{2}\int d{\bf x_1}\int d{\bf x_2}\Delta\rho(\bf {x_1})\Delta\rho({\bf x_2})
{\bar c}^{(0)}({\bf x_1,x_2})\nonumber \\
& & -\frac{1}{2}\int d{\bf x_1} \int d{\bf x_2} \rho({\bf x_1}) \rho({\bf x_2})
{\bar c}^{(n)}({\bf x_1,x_2})
\ea
where 
\ba
\Delta\rho(x)&=&\rho(x)-\rho_l \nonumber \\
&=&\frac{\rho_l}{4\pi}[(1+\Delta\rho^*)\{1+5P_2P_2(\cos{\theta})+9P_4P_4(\cos{\theta})\}-1] \\
{\rm and}\nonumber \\
\Delta\rho^*&=&\frac{\rho_0-\rho_l}{\rho_l} 
\ea 
The minimization of $\Delta W$ with respect to $\rho({\bf x})$ leads to the following relations for 
order parameters 
\ba
P_l&=&\frac{\int_{-1}^{1}P_l(\cos{\theta})\exp(I_1)d\cos{\theta}}{\int_{-1}^{1}\exp(I_1)d\cos{\theta}}\\
1+\Delta\rho^*&=&\frac{1}{2}\int_{-1}^{1}\exp(I_2)d\cos{\theta}
\ea

where 
\be
I_1=\int d{\bf x}_2 \Delta\rho({\bf x}_2)c^{(0)}({\bf x}_1,{\bf x}_2;\rho_0)+
\int d{\bf x}_2 \Delta\rho({\bf x}_2){\tilde c}_{1}^{(n)}({\bf x}_1,{\bf x}_2;\rho)
\ee
and
\be
I_2=\int d{\bf x}_2 \Delta\rho({\bf x}_2)c^{(0)}({\bf x}_1,{\bf x}_2;\rho_0)+
\int d{\bf x}_2 \Delta\rho({\bf x}_2){\tilde c}_{2}^{(n)}({\bf x}_1,{\bf x}_2;\rho)
\ee
The ${\tilde c}_{1}^{n}$ is defined by Eq(3.9) and
\be
{\tilde c_{2}^{n}}({\bf x}_{1},{\bf x}_{2})=2\int_{0}^{1} d\lambda \int_{0}^{1} d\xi \xi 
\int_{0}^{1} d\xi^{'} c^{(n)}({\bf x}_{1},{\bf x}_{2}; \lambda \rho_0;\xi\xi^{'}\sqrt{\sum_{l=1}^n}{P_l}^{2})
\ee

In the isotropic phase the order parameters become zero. Eqs(4.5)-(4.6) are solved self-consistently 
using the values of ${\bar c}^{(0)}$, ${\bar c}^{(n)}$, ${\tilde c}^{(0)}$, ${\tilde c}_{1}^{(n)}$ and
${{\tilde c}_2}^{(n)}$ evaluated in the previous sections. By substituting these solutions
in the expression of $\Delta W$ we locate the transition. At a given temperature and 
density a phase with lowest grand potential is taken as the stable phase. 
Phase coexistence occurs at the value of $\rho_l$ that makes $-\Delta W/N=0$ for the 
nematic and liquid phases. The results are given in Table 1 for both the MSA and PY theories.

\begin{table}[h]
\caption{Isotropic-nematic transition parameters of the model potential with $\beta a_2=1$ and $\beta a_2=0.5$
keeping other parameters fixed at $z_0\sigma=z_2\sigma=1, \beta a_0=0.1$. The pressure is given as $P^*=\beta P/\rho$.}

\begin{tabular}{|c|c|c|c|c|c|c|} \hline
$\beta a_2$&Closure & $\rho_l$ & $\Delta \rho^*$& ${P_2}$ & $P_4$ & $P^*$ \\  \hline
1.0&MSA & 0.490 & 0.120 &0.540&0.150 & 1.480  \\
& &   & & & &\\
1.0&PY&  0.411 & 0.134&0.450 &0.130&1.139  \\
0.5&PY & 0.864 & 0.013& 0.440&0.120&7.700  \\ \hline
\end{tabular}
\end{table}

The pressure can be found using the compressibility equation. In the case of isotropic phase  
\be
\frac{\beta P}{\rho}=1-\frac{1}{\rho}\int_{0}^{\rho}d{\rho^{'}}{\hat c}_{000}(0, \rho^{'})
\ee
For the nematic phase the relation is found to be
\be
\frac{\beta P}{\rho}=\frac{1}{\rho}\int_{0}^{\rho}\frac{d\rho^{'}}{1+\frac{\rho^{'}}{(4\pi)^{3/2}}\int dr r^2 
\sum_{l_1l_2}\sqrt{(2l_1+1)(2l_2+1)}P_{l_1}P_{l_2}h_{l_1l_20}(r,\rho^{'})}
\ee
In Fig 10 we compare the pressure found from the MSA and PY theories for $\beta a_2=1$. In
Fig 11 the pressure found from the PY theory is given for $\beta a_2=0.5$. The plateau corresponds 
to the change in density at the transition. The value of pressure found at the transition is given in
Table 1. 

\section{Discussions} 

We have presented the calculation of the pair correlation functions $h({{\bf x}_1},{{\bf x}_2})$
and $c({{\bf x}_1},{{\bf x}_2})$ in a nematic phase for a model of spherical particles with the long range
anisotropic interaction from the MSA and the PY integral equation theories. 
We chose this model system because for this the OZ equation with the MSA closure has been solved analytically
[14]. The inhomogeneous OZ equation involves the single particle density distribution $\rho({\bf x})$
which we have expressed in terms of order parameters. Non-zero values of order parameters break the 
rotational symmetry. The value of order parameters determine the degree of 
symmetry breaking. For determining the value of order parameters at the isotropic-nematic transition
we used the equation which we found by minimizing the grand thermodynamic potential. The transition 
from the isotropic to nematic in the density-temperature plane is found by solving simultaneously the equations 
for the grand thermodynamic potential and the order parameters. 
This solution gave the value of density, temperature and order parameters at the transition 
which we have listed in Table 1. The value of order parameters in the nematic region has been found 
by minimization of the reduced Helmholtz free energy functional in terms of
order parameters.

The free energy functional given here includes both the symmetry conserving and symmetry broken
parts of the DPCF and therefore correctly describes the ordered phase. In the 
free energy functional of Ramakrishnan and Yousouff[23] the DPCF of the ordered phase is replaced 
by that of the coexisting isotropic fluid.
This amounts to neglecting the symmetry breaking part of the DPCF. In the weighted-density approximation of 
Curtin and Ashcroft and various versions of it[24] the free energy functional is constructed 
in such a way that the free energy density of an inhomogeneous system at a given point is replaced by 
that of a homogeneous system but taken at an auxiliary density which depends parametrically 
on the chosen point. This approach also neglects the new features that emerge in the pair correlation functions due to symmetry breaking. 

One of the important features of the total pair correlation function of
a nematic phase is the appearance of $1/r^*$ tail in harmonic coefficient $h_{l_1l_20m_11-10}$(or
in our notation $h_{l_1l_21}(r^*)$) of $h({\bf x_1},{\bf x_2})$. This has been seen in the 
computer simulation[13] and in the analytical solution of the MSA theory by Holovko and Sokolovska[14].
We found this feature in both the MSA and PY theories. The long-range tail behaviour of $h_{l_ll_21}(r^*)$
is attributed to the director transverse fluctuations which give rise to the Goldstone modes.
This can be seen by taking the tensor order parameter
$Q_{\alpha\beta}=\frac{1}{N}\sum_{i=1}^{N}\frac{3}{2}(e_{i\alpha}e_{i\beta}-\frac{1}{3}
\delta_{\alpha\beta})$ where $\alpha, \beta= x, y, z$ and $e_{i\alpha}$ is the $\alpha$ component
of the molecular axis vector ${\bf e_i}$ of each molecule and $\delta_{\alpha \beta}$ the Kronecker
symbol and calculating (assuming that the director is along $z$ and the $y-$axis is perpendicular
to wave vector ${\bf k}$) the correlation $\langle Q_{xz}(k) Q_{xz}(-k)\rangle$. The result involves coefficients
$h_{l_1l_2lm_1m_2m}^{(n)}(k)$ with $|m_1|$, $|m_2|$=1. These coefficients which are the
Fourier transform of $h_{l_1l_2lm_1m_2m}^{(n)}(r^*)$ behave as $1/k^2$ for
$k\to 0$. 

The calculation of $A_{ex}^{(n)}[\rho]$ involves integration over $c^{(n)}({\bf x_1},{\bf x_2};[\rho])$
in the density space. This space is characterized by two variables $\lambda$ and $\xi$ which vary from 0 to 1
and raise the density from 0 to $\rho_0$ (the average number density of the ordered phase) and order parameters from 
0 to their final values. One has therefore to evaluate $c^{(n)}({\bf x_1},{\bf x_2}; [\rho])$ from 
zero value of number density and order parameters to their final values at small intervals.
This may need large computational investment. We have therefore examined the possibility of  
calculating $A^{(n)}[\rho]$ by considering the DPCF which involves only the principal order parameter 
i.e. $P_2$. The resulting 
free energy functional has been found to give (see Sec III) results which are close 
to the exact results. 

It is important to note that the density-functional approach allows one to include more order 
parameters in the theory even though they are not included in calculating $A_{ex}^{(n)}[\rho]$. 
This is done through the parametrization
of $\rho({\bf x})$[21].  The results given in Table 1 and in Sec III for the PY theory correspond to 
this approximation. The harmonic coefficients plotted in Figs 1-9 have been calculated using both $P_2$
and $P_4$.

The harmonic coefficient $h_{l_1l_21}(r^*)$ has been found to be sensitive to values of $P_2$ and $P_4$.
While $P_2$ creates the $1/r^*$ tail (or equivalently makes its Fourier transform to diverge at 
k=0) $P_4$ suppresses it. This can be seen from the OZ expression
\be
{\hat h}_{221}(k)=\frac{{\hat c}_{221}(k)}{1+\frac{\rho^*}{(4\pi)^{3/2}}[1+0.714 P_2
-1.714 P_4]{\hat c}_{221}(k)}
\ee

When at $k\to 0$ the second term in the denominator approaches to -1 the divergence occurs.
Since ${\hat c}_{221}(k\to 0)$ is negative the term involving $P_2$ help  
while the term involving $P_4$ opposes the divergence.

The theory developed here can be extended to other ordered phases. Since the symmetry breaking 
part of pair correlation functions have features of the ordered phase including its geometrical packing,
the free energy functional described here will allow us to study various phenomena of ordered phases.
Our work on freezing of simple liquids into crystalline solids is in progress and the results will be 
reported in near future.

\section{\bf Acknowledgment:}
This work was supported by a research grant from DST of Government of India, New Delhi. One of us (P. M.)
would like to thank Prof. T. V. Ramakrishnan for his support and JNCASR (Bangalore) for research fellowship.



\begin{thebibliography}{99}
\bibitem{1}  G. R. Luckhurst and P. S. J Simmonds, Mol. Phys.  {\bf 80}, 233 (1993);
             M. A. Bates and G. R. Luckhurst, J. Chem. Phys. {\bf 110}, 7087 (1999). 
\bibitem{2}  E. de Miguel, L. F. Rull, M. K. Chalam, K. E. Gubbins
             and E. V. Swol, Mol. Phys. {\bf 72}, 593 (1991); E. de Miguel, L. F. Rull, M. K. Chalam and K. E. Gubbins,
             Mol. Phys. {\bf 74}, 405 (1991); E. de Miguel, E. Martin del Rio, J. T.Brown and M. P. Allen,
             J. Chem. Phys. {\bf 105}, 4234 (1996); J. T. Brown, M. P. Allen and E. Martin del Rio, and E. de Miguel,
             Phys. Rev. E {\bf 57}, 6685 (1998); E. de Miguel, Mol. Phys {\bf 100}, 2449 (2002); 
             E. de Miguel and E. Martin del Rio, J. Chem. Phys. {\bf 118}, 1852 (2003). 
\bibitem{3}  M. P. Allen, J. T. Brown and M. A. Warren, J. Phys.: Condens. Matter
             {\bf 8}, 9433 (1996).
\bibitem{4}  L. Longa, G. Cholewiak, R. Terbin and G. R. Luckhurst,
              Eur. Phys. J. E {\bf 4}, 51 (2001).
\bibitem{5} N. H. Phoung, G. Germano and F. Schmid, J. Chem. Phys.{\bf 115}, 7227(2001);
             N. H. Phoung, G. Germano and F. Schmid, Comput. Phys. Commun. {\bf 147}, 350 (2002).
\bibitem{6} J. P. Hansen and I. R. McDonald, {\it Theory of Simple Liquids}(Academic, London, 1986), 2nd ed;
             C. G. Gray and K. E. Gubbins, {\it Theory of Molecular Fluids}(Oxford, New York, 1984), Vol I.
\bibitem{7} J. Ram, R. C. Singh and Y. Singh, Phys. Rev. E {\bf 49}, 5117 (1994);
             R. C. Singh, J. Ram and Y. Singh, Phys. Rev. A  {\bf 54}, 977 (1996);
             R. C. Singh, J. Ram and Y. Singh, Phys. Rev. E {\bf 65}, 031711(2002);
             P. Mishra, J. Ram and Y. Singh, J. Phys.: Condens. Matter {\bf 16}, 1695 (2004);
             P. Mishra and J. Ram, Eur. Phys. J. E {\bf 17}, 345 (2005).
\bibitem{8} M. Letz and A. Latz, Phys. Rev. E {\bf 60}, 5865 (1999). 
\bibitem{9} A. Yethiraj and G. Stell, J. Stat. Phys. {\bf 100}, 39(2000). 
\bibitem{10} A. Perera, P. G. Kausalik, and G. N. Patey, J. Chem. Phys. {\bf 87}, 1295 (1987). 
\bibitem{11} D. L. Cheung, L. Anton, M. P. Allen, and A. J. Masters, Phys. Rev. E {\bf 73}, 061204 (2006).
\bibitem{12} J. S. McCarley and N. W. Ashcroft, Phys. Rev. E {\bf 55}, 4990 (1997).
\bibitem{13} N. H. Phoung and F. Schmid, J. Chem. Phys. {\bf 119}, 1214 (2003);
\bibitem{14} M. F. Holovko and T. G. Sokolovska, J. Mol. Liq. {\bf 82}, 161 (1999).   
\bibitem{15} T. G. Sokolovska, R. O. Sokolovskii and M. F. Holovko, Phys. Rev. E {\bf 62}, 6771 (2000).   
\bibitem{16} P. G. de Gennes and J Prost, {\it The Physics of Liquid
             Crystals} (Clarendon, Oxford, 1993), 2nd ed.
\bibitem{17} R. A. Lovett, C. Y. Mou, E. P. Buff, J. Chem. Phys. {\bf 65}, 570 (1976);
            M. S. Wertheim, J. Chem. Phys. {\bf 65}, 2377 (1976).
\bibitem{18}H. Zhong and R. G. Petschek, Phys. Rev. E {\bf 51}, 2263 (1994).
\bibitem{19} P. Mishra and Y. Singh, Phys. Rev. Lett. {\bf 97}, 177801 (2006).
\bibitem{20}  J. G. Gay and B. J. Berne, J. Chem Phys. {\bf 74}, 3316 (1981);
             J. Chem. Phys. {\bf 105}, 4234 (1996).
\bibitem{21} Y.Singh, Phys.Rep. {\bf 207}, 351 (1991).
\bibitem{22} I. Paci and N. M. Cann, J. Chem. Phys. {\bf 119}, 2638 (2003);
             S. H. L. Klapp and G. N. Patey, J. Chem. Phys. {\bf 112}, 3832 (2000);
             L. Blum and A. J. Torruella, J. Chem. Phys. {\bf 56}, 303 (1972).
\bibitem{23} T. V. Ramakrishnan and M. Yussouff, Phys. Rev. B {\bf 19}, 2775 (1979); A. D. J. Haymet
            and D. Oxtoby, J. Chem. Phys. {\bf 74}, 2559(1981).
\bibitem{24} W. A. Curtin and N. W. Ashcroft, Phys. Rev. A {\bf 32}, 2909 (1985); 
             P. Tarazona, Phys. Rev. A {\bf 31}, 2672 (1985); Phys. Rev. A {\bf 32},
             3148(E)(1985); A. R. Denton and N. W. Ashcroft, Phys. Rev. A {\bf 39}, 4701 (1989); 
             M. Baus, J. Phys. Condens. Matter {\bf 1}, 3131(1989); 
             J. F. Lutsko and M. Baus, Phys. Rev. Lett. {\bf 64}, 761(1990). 
\bibitem{25}J. P. Hansen and L. Verlet, Phys. Rev. {\bf 184}, 150 (1969).
\bibitem{26} W. F. Saam and C. Ebner, Phys. Rev. A {\bf 15}, 2566 (1977).
\end{thebibliography}
\end{document}